\documentclass[10pt,twocolumn,twoside]{IEEEtran}
\usepackage{graphicx,cite}
\usepackage{array}

\usepackage{color}

\usepackage{ifthen}

\usepackage{setspace}
\usepackage{amsmath,amsfonts,amssymb}

\newcommand{\RemoveThis}{\ifthenelse{1< 0}}

\newcommand{\boldeta}{\boldsymbol{\eta}}
\newcommand{\boldlambda}{{\boldsymbol{\lambda}}}
\newcommand{\boldmu}{\boldsymbol{\mu}}

\newcommand{\boldnu}{\boldsymbol{\nu}}
\newcommand{\lambdabar}{{\overline{\lambda}}}
\newcommand{\mubar}{{\overline{\mu}}}
\newcommand{\nubar}{{\overline{\nu}}}

\newcommand{\Reals}{{\mathbb{R}}}

\newtheorem{lemma}{\textbf{Lemma}}

\newtheorem{theorem}{\textbf{Theorem}}

\newtheorem{corollary}{\textbf{Corollary}}

\begin{document}

\title{Pigouvian Tolls and Welfare Optimality with Parallel Servers and Heterogeneous Customers}

%

\author{
  \begin{tabular}{ccc}
    Tejas Bodas      & Ayalvadi Ganesh            & D.~Manjunath \\
    IIT Dharwad, INDIA & University of Bristol, UK  & IIT Bombay, INDIA 
  \end{tabular}
}


\maketitle

\begin{abstract}
Congestion externalities are a well-known phenomenon in transportation and 
communication networks, healthcare etc. Optimization 
by self-interested agents in such settings typically results in equilibria which are
sub-optimal for social welfare. Pigouvian taxes or tolls, which impose a user 
charge equal to the negative externality caused by the marginal user to other 
users, are a mechanism for combating this problem. 
In this paper, we study a non-atomic congestion game in which heterogeneous 
agents choose amongst a finite set of heterogeneous servers. The delay at a 
server is an increasing function of its load. Agents differ in their sensitivity to delay. 
We show that, while selfish optimisation by agents is sub-optimal for social welfare, 
imposing admission charges at the servers equal to the Pigouvian tax causes 
the user equilibrium to maximize social welfare. In addition, we characterize the 
structure of welfare optimal and of equilibrium allocations.

\end{abstract}

\section{Introduction}
\label{sec:intro}

We study service systems in which customers or agents can be served by any one 
of several heterogeneous servers. Customers arrive into the system according to a 
random process, reside in the system while being served, and then depart. Customers 
differ in their aversion to some congestion-based metric such as their sojourn-time in 
the system or the number of other customers with whom they share the server. We 
seek to determine how customers may be assigned to servers in such a way as to 
optimize some social welfare function, and also how pricing may be used to incentivize 
selfish customers to achieve the same social optimum.

Examples of such systems include web server farms, cloud and grid computing clusters, 
communication networks and cognitive radio systems. In these examples, customers 
may differ in the quality of service they require, and in their willingness to pay for it. 
The quality of service of a customer may depend on the share of bandwidth or other 
resources it receives, or the service latency or the sojourn time in the system. Another 
example arises in transportation, where users may have a choice of tolled and toll-free 
routes, or between multiple modes of transport. Further examples include healthcare, 
where patients may be choosing between different service providers. Our modeling 
framework is quite general in this regard and encompasses all the above examples. 

A common feature of the above examples is that the more customers choose a 
particular server, the worse their individual experience. For example, if more drivers 
choose a certain road, the slower the flow of traffic on it (above a certain utilization) 
and hence the longer the journey time. Similarly, if more patients choose a certain 
hospital, then they may have to wait longer for treatment, at least in the short run, 
when service capacities cannot be changed. This is known as a {\bf congestion externality.}

Customer preferences are captured by a cost function that could depend on the 
system occupancy or sojourn time in a fairly arbitrary way. For example, in a 
transportation network, the cost function could be the expectation of a given function 
of the travel time, e.g., the probability that the travel time exceeds a certain threshold 
value. In a communication network, it could be a function of the bandwidth received, 
or the latency, or a combination of the two. We allow for customer heterogeneity 
by applying a suitable multiplier to the congestion cost. We call this multiplier its 
{\bf delay-sensitivity} (but emphasise that congestion costs can take account of factors 
other than delay). 

We do not constrain service policies except to insist that they be non-discriminatory 
and agnostic of customer characteristics. Thus, for instance, one server may adopt 
a first-come first-served (FCFS) policy while another splits its capacity equally amongst 
all its customers (processor-sharing or PS). Servers may charge a fixed admission price 
to each customer choosing that server; these can be different between servers but 
must be the same for each customer. In particular, servers cannot charge for priority 
or preferential treatment.

Customers choose a server so as to optimize their individual expected utility, i.e., 
to minimize the sum of the admission price and the expected congestion cost 
(weighted by their own delay-sensitivity). As the congestion cost depends on the 
choices of other customers, the interaction between them constitutes a game. 
The payoff structure makes this a {\bf congestion game}~\cite{Rosenthal73, 
Monderer96}. We assume in addition that customers are infinitesimal, i.e., 
that the impact of a marginal customer on the congestion cost at any server 
is negligible. This assumption renders the congestion game non-atomic. 
Nash equilibria in non-atomic congestion games are also known as Wardrop 
equilibria, from their origins in transportation networks~\cite{Wardrop52};
see~\cite[Chapter 18]{Nisan07} for an overview of congestion games. 

The goal of this paper is to study the social cost, i.e., the sum of congestion costs 
incurred by different customers weighted by their sensitivity to congestion, of a 
Wardrop equilibrium. In particular, we want to know if admission prices can be set 
in such a way as to ensure that the social cost at equilibrium is the minimum 
achievable by a central planner who could assign customers to servers. We answer 
this question in the affirmative. One set of such prices admit an interpretation as 
Pigouvian taxes associated with congestion externalities at the servers. While the 
welfare optimality of Pigouvian taxes is known in general, our contribution in this paper 
is to show that these depend only on the server, and not on the customer type. In 
other words, all customers using the same server are charged the same levy (which 
may depend on the mix of customer types choosing that particular server).

A second contribution of the paper is a characterisation of the structure of socially 
optimal allocations and of Wardrop equilibria. Specifically, we show that in an optimal 
allocation, the server with the smallest congestion cost serves the most delay-sensitive 
customers, the one with the next smallest congestion cost serves the next most 
sensitive set of customers, and so on. We show that, for arbitrary admission prices 
at the servers, Wardrop equilibria have the same structure. Furthermore, the higher 
the admission price at a server, the lower its congestion cost (among servers that 
are utilized by some customer).

We survey some related work in the remainder of this section, before presenting 
a formal statement of the model and problem in the next, and stating our main 
results. Proofs are presented in the following section, and we conclude with a 
discussion of limitations of the current work and some open problems.

\subsection{Related Work}
\label{sec:prev-work}

The notion of a congestion externality was first formalized by Pigou~\cite{Pigou20}, 
who proposed the use of a charge or levy to internalize the congestion externality 
in transportation networks, thereby guiding the system to a social optimum. Such 
charges are known as Pigouvian taxes and have since been studied in a wide variety 
of contexts including queueing systems \cite{Littlechild74,Bradford96}, transportation 
networks~\cite{Smith79a,Yang98}, matching markets~\cite{He18} and 
climate change~\cite{Mankiw09}. While much of the work on Pigouvian taxes 
focuses on achieving socially optimal levels of consumption of a good associated 
with externalities, the work in this paper is most relevant when demand is inelastic 
(i.e., the quantity of demand does not depend on the price), 
but there is a choice between substitutes which generate different externalities. 
This is the case in many queueing and transportation applications. Secondly, our 
work considers heterogeneous agents, with different delay-sensitivites. In the 
following, we refer to them as multiclass customers, with ``class" being used 
as a synonym for ``delay-sensitivity".


There is a substantial literature on the allocation of multi-class customers 
to parallel queues in both centralized and decentralized settings, including 
a variety of pricing schemes and game-theoretic formulations. Much of 
this work looks at specific cost functions arising from those models, 
whereas we consider a more general and abstract formulation. Below, we 
describe some of the work more closely related to the approach taken in 
this paper and delineate these from the results we present. We use Kendall's 
notation for queueing models, which we now briefly describe. A queue is 
described by a triple $X/Y/n$, where $X$ describes the arrival process, 
$Y$ the job size distribution, and $n$ the number of servers. Common 
choices for $X$ are $M$, denoting Markovian and referring to a Poisson 
arrival process, and $G$, denoting ``general" and referring to an arrival 
process in which the inter-arrival times are independent and identically 
distributed (i.i.d.), but with a general distribution. (Some authors prefer $GI$ 
to emphasise the assumption of independence.) Common choices for $Y$ 
are $M$, denoting Markovian and referring to job sizes with an exponential 
distribution, $G$, denoting i.i.d. job sizes with a general distribution, and $D$, 
denoting fixed, deterministic job sizes. If the service discipline is not the 
default $FCFS$ discipline, it is added to the notation. Thus, for example, 
an $M/G/1-LCFS$ queue has Poisson arrivals, i.i.d. job sizes with a general 
distribution, and a single server which adopts a last-come-first-served policy.

There are several works that study the use admission prices to reduce congestion.
Naor~\cite{Naor69}, Edelson and Hilderbrand~\cite{Edelson75} and 
Littlechild~\cite{Littlechild74} studied $M/M/1$ queues with identical customers 
who must choose between paying an admission price to enter the queue, 
incurring a random delay and receiving a fixed reward for service, or balking 
(i.e., leaving without being served). Admission prices are set by an operator who 
seeks to maximize revenue. If customers can observe the queue length on 
arrival and base their balking decision on it, then the revenue-maximizing 
admission price exceeds the one that maximizes social welfare~\cite{Naor69}. 
However, if customers cannot observe the queue but must base their decision 
on only the known arrival and service rates, then these two admission prices 
coincide~\cite{Edelson75,Littlechild74}. In the latter setting, Littlechild~\cite{Littlechild74} 
obtained the admission fee as a Pigouvian tax and showed that this will induce a 
socially optimal arrival rate. Bradford~\cite{Bradford96} extended the results to 
multiclass customers, each with their own delay cost function and reward for service, 
and obtains the Pigouvian admission charge for each class that achieves the socially
optimal allocation. The admission charge is independent of the queue from which the 
customer receives service but depends on its class, which means that the system 
needs to elicit information of the customer class. In contrast, admission charges in 
our model are calculated for each queue but are agnostic of the customer class. 

The equilibrium allocation of customers in multiqueue systems was studied by Bell 
and Stidham~\cite{Bell83}, and Haviv and Roughgarden~\cite{Haviv07}. Both 
works focused on homogeneous customers, i.e., a single customer class. Bell 
and Stidham~\cite{Bell83} studied a set of parallel $M/G/1$ queues which differ 
in their holding cost per unit time and in their mean service time. They 
established structural properties of a socially optimal allocation as well as of 
Wardrop equilibria. Restricting their attention to parallel $M/M/1$ queues, 
Haviv and Roughgarden~\cite{Haviv07} obtained an upper bound on the price 
of anarchy (PoA), defined as the ratio of the total cost at the Wardrop equilibrium 
to that at the social optimum. In comparison, we consider multiclass customer 
populations and general cost functions.

Borst~\cite{Borst95} studied the probabilistic allocation of multiclass traffic to parallel 
$M/G/1$ queues so as to minimize a specific social cost function, namely the total 
mean waiting cost per unit of time. He established a structural property of the optimal 
allocation. The structure we obtain for the optimal allocation is essentially the same, 
but our results apply to a very general class of queueing models and cost functions; 
we also do not restrict to finitely many customer classes. In addition, we consider a 
game-theoretic setting of selfish optimization and determine a pricing mechanism that 
will achieve social optimality with selfish optimization. 

Sethuraman and Squillante~\cite{Sethuraman99} considered a variant of 
this problem where, in addition to optimal routing, servers decide the order 
in which customers in a queue are served, depending on their class, so as 
to optimise social welfare. An alternative approach is to allow customers to 
purchase priorities~\cite{Balachandran72,Lui85,Rao98,Afeche04,Kittsteiner05}; 
a comprehensive survey of these and other similar models is presented by 
Hassin and Haviv~\cite{Hassin03}. Our work differs in that we do not allow 
servers to discriminate between customers, as a consequence of which they 
do not need to elicit information about customer class. This may be more 
realistic in certain applications.

A number of works have studied specific applications in which pricing is used 
to achieve service differentiation by incentivising end users to segregate 
themselves on the basis of their willingness to pay for higher quality or lower 
delay. In particular, there is a substantial body of work proposing charging for 
differentiated services (Diffserv) in the Internet, and studying the resulting 
user strategies and equilibria; see \cite{Odlyzko99,Jain01,Dube02,Borkar04}, 
for example. Additional examples include queues~\cite{Tandra04} and 
transport networks~\cite{Yang04}. There has also been work on models 
in which prices are dynamically adapted in response to observed 
demands~\cite{Ganesh07}; it is shown that if prices adapt sufficiently slowly, 
then the system converges to a Nash equilibrium. Finally, while the work 
presented in this paper focuses on parallel queues, there has been considerable 
work on general networks; see Roughgarden~\cite{Roughgarden05} for a 
detailed discussion of selfish routing and the PoA, and Fleischer 
\emph{et al.}~\cite{Fleischer04} for the analysis of equilibria in a very general 
network model.

\section{Model and Results}
\label{sec:prelims}

Consider a system with $N$ parallel channels for service, which we refer to as servers or 
queues. Customers arrive into the system according to a marked Poisson process with 
intensity $\eta \times F$; here, $\eta$ denotes the arrival rate, and $F$ the distribution 
of the arriving customer's class or delay-sensitivity. The only assumption we make about 
the distribution $F$ is that its support is bounded away from zero and infinity, i.e., that 
there are constants $\beta_{\min}>0$ and $\beta_{\max}<\infty$ such that $F(x)=0$ 
for all $x<\beta_{\min}$, and $F(\beta_{\max})=1$. Arriving customers must either select 
or be allocated to one of the queues upon arrival. We assume that the allocation has to be 
made with no knowledge of current or past queue occupancies, or past arrival times or 
routing decisions. Such an assumption may be less realistic for centralized allocation than 
when customers make individual decisions. Nevertheless, imposing this assumption uniformly 
permits clearer comparison of the two settings. The structure of Wardrop equilibria can be 
very different if queue occupancies are known, and requires a separate analysis, which is a 
topic for future research. In general, providing additional information can make the Wardrop 
equilibrium worse for all agents~\cite{Acemoglu18}!

Under the assumption that queue occupancies are unknown, it is natural to restrict 
attention to Markovian policies, which route customers to queues according to some 
fixed probability vector that may depend on the customer's class, but not on history. 
(If queue occupancies are known, policies are Markovian with respect to a larger 
state space which includes that information.)
We assume that customers of all classes have the same job size distributions, 
and that, once they join a queue, they are treated identically within it. Consequently, 
we assume that the congestion cost associated with a queue depends only on the 
aggregate arrival rate into that queue (and its service capacity and policies), but not 
on the composition of those arrivals. We make this precise below.

Let $\boldeta$ denote the Borel measure on $[0,\beta_{\max}] \subset \Reals_+$ 
defined on intervals by 
\begin{equation} \label{eq:etadef}
\boldeta((a,b]) = \eta(F(b)-F(a)).
\end{equation}
In other words, the measure of an interval $(a,b]$ is defined as the total arrival rate 
of customers whose class lies in this interval. As usual, the measures of all Borel sets 
are determined by those of intervals. All measures in this paper are non-negative, 
finite Borel measures.

Now, Markovian routing corresponds to a decomposition of the measure $\boldeta$ as 
\begin{equation} \label{eq:lambdadef}
\boldeta = \boldlambda_1+\ldots+\boldlambda_N,
\end{equation} 
where $\boldlambda_j$ is a measure on $[\beta_{\min},\beta_{\max}]$ for each 
$j=1,\ldots,N$; arrivals into the $j^{\rm th}$ queue of customers with classes in 
$(a,b]$ constitute a Poisson process of rate $\boldlambda_j((a,b])$. We denote 
the total arrival rate into the $j^{\rm th}$ queue, and the mean delay-senstivity 
of arrivals into this queue, by 
\begin{equation} \label{eq:meandefs}
\lambda_j= \boldlambda_j([\beta_{\min},\beta_{\max}]) \mbox{ and } 
\lambdabar_j = \int_{\beta_{\min}}^{\beta_{\max}} \beta d\boldlambda_j(\beta),
\end{equation} 
respectively.

Next, we associate with each queue $j$ a cost function $D_j(\cdot)$ which specifies 
the congestion cost generated by a given aggregate arrival rate; thus, $D_j(\lambda)$ 
is the congestion cost incurred by each customer when the arrival rate into queue $j$ 
is $\lambda$. The cost could be the mean sojourn time, or some higher moment of it, 
or the probability of the sojourn time exceeding a specified threshold. Our only 
assumption is that each function $D_j$ be monotone increasing, continuous, and 
continuously differentiable in the interior of its domain (the set of arrival rates for which 
$D_j$ is finite), with strictly positive derivative. In particular, we assume that the 
domain of each $D_j$ is either $\Reals_+$ or an interval of the form $[0,a)$, and 
that in the latter case, $\lim_{x\uparrow a} D_j(x) = +\infty$.


The assumptions above are rather mild. We do not restrict the number of servers 
at a queue or the service discipline. Indeed, different queues may have different 
numbers of servers and employ different service disciplines. They can also be 
associated with different cost functions, for example the mean sojourn time at 
one queue and the second moment at another. The only requirement is that 
each queue treat all customers alike, irrespective of their class. In addition to 
traditional queueing models, our set-up also encompasses transportation models, 
where the mean journey time on a road may be some increasing function of the traffic 
intensity on it. The main motivation for the assumption of Poisson arrivals is that 
it makes each $D_j$ a function of a single real variable. It is not obvious how the 
monotonicity and differentiability assumptions would generalize if $D_j$ were to be 
a function of the law of a stochastic process.

We are now ready to state the social welfare maximization problem.
The objective is
\begin{equation} \label{welfare_opt_prob}
\begin{aligned}
\inf_{\boldlambda_1,\ldots,\boldlambda_N} & \mathcal{U}(\boldlambda_1,\ldots,\boldlambda_N) 
= \sum_{j=1}^N \lambdabar_j D_j(\lambda_j), \\
\mbox{subject to } & \boldlambda_1 + \ldots + \boldlambda_N = \boldeta.
\end{aligned}
\end{equation}
Thus, the social cost is defined as the sum of the expected costs incurred 
by customers of different classes at different queues, weighted by the 
corresponding flow rates.

Our first result states that, if the social cost minimization problem is feasible, then it 
has a solution, i.e., the minimum is attained. 
\begin{lemma} \label{lem:soc_opt}
Let $\boldeta$ be a finite measure with bounded support. Suppose that the cost 
functions $D_j$, $j=1,\ldots,N$, satisfy the assumptions stated above. If the 
optimization problem in \eqref{welfare_opt_prob} is feasible, i.e., there is some 
decomposition $(\boldlambda_1,\ldots,\boldlambda_N)$ of $\boldeta$ such that 
$D_j(\lambda_j)$ is finite for all $j=1,\ldots,N$, then \eqref{welfare_opt_prob} 
has a solution $(\boldlambda_1^*,\ldots,\boldlambda_N^*)$.
\end{lemma}

Next, we consider the formulation of a game between customers. Here, we 
allow the queues to charge admission prices, denoted by $c_j$ at queue $j.$ 
The goal of a class $\beta$ customer entering the system is to choose a 
queue $j$ so as to minimize $c_j+\beta D_j(\lambda_j)$ where $\lambda_j$ 
is determined through the strategies of all customers. We assume that the 
arrival intensity measure $\boldeta$ and the cost functions $D_j(\cdot)$, 
$j=1,\ldots,N$ are common knowledge. As we assumed that customers 
do not have access to current or past queue occupancies, or the history 
of arrival times or routing choices, they are necessarily restricted to choosing 
a server according to a fixed probability distribution, albeit one that may 
depend on their class. Thus, once again, the joint strategies may be 
represented by a decomposition of the measure $\boldeta$ into measures 
$\boldlambda_1,\ldots,\boldlambda_N$. We want to know when such a 
decomposition corresponds to a Wardrop equilibrium of the game. 

The condition for a decomposition $(\boldlambda_1,\ldots,\boldlambda_N)$ 
of $\boldeta$ to be a Wardrop equilibrium is that  
\begin{equation} 
\label{wardrop}
\begin{aligned}
&c_j+\beta D_j(\lambda_j) \leq c_k+\beta D_k(\lambda_k) \\
&\forall \;
j,k =1,\ldots,N, \mbox{ and } \beta \in \mbox{supp}(\boldlambda_j),
\end{aligned}
\end{equation}
%
where $\mbox{supp}(\boldeta)$ denotes the support of the measure $\boldeta$, 
namely the smallest closed set $F$ such that $\boldeta(F^c)=0$. Here, $F^c$ 
denotes the complement of $F$. The condition in \eqref{wardrop} roughly says that, 
if a positive mass of customers of class $\beta$, or in an arbitrarily small neighbourhood 
of it, use queue $j$, then the expected cost of a class $\beta$ customer in that queue 
must be no higher than its expected cost in any other queue. 

The existence of a Wardrop equilibrium can be shown by looking at an auxiliary 
optimization problem, following Beckmann \emph{et al.}~\cite{Beckmann56} in 
the single-class setting, and Yang and Huang~\cite{Yang04} in the multiclass 
setting with a finite number of classes. Consider the optimization problem 
\begin{equation} \label{eq:wardrop_opt}
\begin{aligned}
&\inf \; \hat{\mathcal{U}} (\boldlambda_1\,\ldots,\boldlambda_N) \\
&\qquad = 
\sum_{j=1}^N \Bigl( \int_0^{\lambda_j} D_j(x)dx  + 
c_j \int_{\beta_{\min}}^{\beta_{\max}} 
\frac{1}{\alpha}d\boldlambda_j(\alpha) \Bigr), \\
&\mbox{subject to } \boldlambda_1 + \ldots + \boldlambda_N = \boldeta.
\end{aligned}
\end{equation}
The existence of a solution follows by Lemma~\ref{lem:soc_opt}. It can easily 
be shown that any solution satisfies \eqref{wardrop}, which are essentially 
first-order conditions for optimality in the auxiliary problem. We include a formal 
statement and proof for completeness.

\begin{lemma} \label{lem:wardrop_exist}
The infimum in the optimization problem \eqref{eq:wardrop_opt} is attained. Moreover, 
any minimizer $(\boldlambda^W_1,\ldots,\boldlambda^W_N)$ is a Wardrop equilibrium, 
i.e., it satisfies the condition in \eqref{wardrop}.
\end{lemma}

A natural mechanism design\footnote{Mechanism design deals with the problem 
of achieving desired social choice objectives by designing the rules of the game 
such that the socially desirable outcome is a Nash equilibrium of the game; 
see~\cite{Nisan07} for further details.}  
question is whether we can set admission prices in such a way that selfish users 
reacting to these prices would assign themselves to queues in the proportions 
required for optimizing social welfare. Our main result affirms that this is indeed 
the case if admission prices are set equal to Pigouvian taxes corresponding to a 
welfare-optimal allocation.
\begin{theorem}
\label{thm:prices_optimal}
Let $(\boldlambda^*_1,\ldots,\boldlambda^*_N)$ be a solution of the social 
cost minimization problem, \eqref{welfare_opt_prob}. Set the admission price 
$c_j$ at queue $j$ to be 
\begin{equation} \label{eq:pigou_tax}
c_j = \lambdabar_j D'_j(\lambda^*_j),
\end{equation} 
where $D'_j$ denotes the derivative of $D_j$.

Then, $(\boldlambda^*_1,\ldots,\boldlambda^*_N)$ is a Wardrop equilibrium, 
i.e., it satisfies \eqref{wardrop} with these admission prices.
\end{theorem}
Notice that $c_j$ given in \eqref{eq:pigou_tax} is precisely the total negative 
externality imposed on existing customers at this queue by the admission 
of a marginal customer, and is hence the Pigouvian toll for this queue.

We now turn to the question of computing the optimal decomposition 
of a given measure $\boldeta$. If we can compute the optimal allocation, 
then we can also compute the corresponding Pigouvian taxes. Note that 
we start by assuming that the measure $\boldeta$ is given. In practice, 
one of the major challenges of implementing Pigouvian taxes is eliciting 
utility functions; in our context, that corresponds to eliciting the true 
delay sensitivities $\beta$ of different agents. Getting agents to truthfully 
reveal their preferences is a major challenge in mechanism design, and 
one which we do not address in this paper. Instead, we restrict ourselves 
to computing the optimal allocation \emph{given} the true distribution of 
delay sensitivities. 

The constraint on $(\boldlambda^*_1,\ldots,\boldlambda^*_N)$  in the 
optimization problem \eqref{welfare_opt_prob} is linear, and so the set 
of measures satisfying the constraint is convex. If the cost function 
$\sum_{j=1}^N \lambdabar_j D_j(\lambda_j)$ were a convex function 
of $(\boldlambda^*_1,\ldots,\boldlambda^*_N)$, then the optimization 
problem would be convex, and could be solved using gradient descent 
methods. Unfortunately, this is not necessarily the case, as illustrated by 
the following counterexample.

Consider a system with two classes of customers and two $M/M/1$ queues. 
Class~$i$ customers arrive according to a Poisson process of rate $\eta_i$ 
and have delay sensitivity $\beta_i$. Thus, the arrival intensity measure is 
$\boldeta=\eta_1 \delta_{\beta_1} + \eta_2 \delta_{\beta_2}$, where 
$\delta_x$ denotes the Dirac delta which puts unit mass at $x$. The job sizes 
for both classes are assumed to be i.i.d. exponential random variables with unit 
mean. Both servers have a unit service rate. We assume that $\eta_1+\eta_2 
< 1$, so that all allocations are feasible. 

Recall that the mean delay in an $M/M/1$ queue with arrival rate $\lambda$ 
and service rate 1 is $1/(1-\lambda)$. Hence, the (class-weighted) congestion 
cost corresponding to a decomposition $(\boldlambda_1,\boldlambda_2)$ of 
$\boldeta$ is given by 
$$
\mathcal{U}(\boldlambda_1,\boldlambda_2) = \frac{\lambdabar_1}{1-\lambda_1} 
+ \frac{\lambdabar_2}{1-\lambda_2}.
$$
The constraint that $\boldlambda_1$ and $\boldlambda_2$ are non-negative 
and decompose $\boldeta$ is equivalent to the constraints that $\lambda_1 
+ \lambda_2 = \eta_1+\eta_2$, $\lambdabar_1+\lambdabar_2 = 
\beta_1 \eta_1 + \beta_2 \eta_2$, and that they are all non-negative. 
Thus, the welfare optimization problem \eqref{welfare_opt_prob} can be 
rewritten as
\begin{equation} \label{welfare_opt_2class}
\begin{aligned}
\inf & \; U(\lambda_1, \lambdabar_1, \lambda_2, \lambdabar_2)  = 
\frac{\lambdabar_1}{1-\lambda_1} + \frac{\lambdabar_2}{1-\lambda_2}, \\
\mbox{subject to} &  \; \lambda_1+\lambda_2 = \eta_1+\eta_2, \\ 
& \; \lambdabar_1+\lambdabar_2 = \beta_1 \eta_1 + \beta_2 \eta_2, \\
& \; \lambda_1, \lambdabar_1, \lambda_2, \lambdabar_2 \geq 0.
\end{aligned}
\end{equation}
We now have the following negative result.
\begin{lemma} \label{lem:nonconvex}
The optimization problem in \eqref{welfare_opt_2class} is not convex.
\end{lemma}

In view of the above lemma, it is not obvious how to numerically compute 
socially optimal allocations in general. Nevertheless, we show below that both 
socially optimal allocations and Wardrop equilibria possess nice structural 
properties. These might suggest efficient algorithms for finding optima and 
equilibria in the model studied here.

%
%

\begin{theorem} 
\label{thm:welfare-opt-cts} 
Let $(\boldlambda_1^*,\ldots,\boldlambda_N^*)$ achieve the minimum in 
\eqref{welfare_opt_prob}.  Suppose $i$ and $j$ are distinct queues, $\beta_2 
> \beta_1 \geq 0$, and
$$
\boldlambda^*_j([\beta_2,\infty))>0 \mbox{ and } \boldlambda^*_i([0,\beta_1])> 0.
$$
Then $D_{i}(\lambda^*_i) > D_{j}(\lambda^*_j).$ This inequality also holds if 
$\lambda^*_i=0$ and $\lambda^*_j>0$.
\end{theorem}

The theorem says that if some of the customers served at queue $j$ have higher 
delay sensitivity than some of the customers served at queue $i$ (where ``some" 
is to be interpreted as ``a set of positive measure"), then the congestion cost at 
queue $j$ must be smaller. Moreover, any queue which serves no customers (or 
a set of measure zero) must have larger congestion cost than any queue which 
serves some customers. The theorem implies that the queues segregate traffic 
by class as follows:
 
\begin{corollary} 
\label{cor:sorting-cts}
Suppose $(\boldlambda_1^*,\ldots,\boldlambda^*_N)$ solves the optimization 
problem \eqref{welfare_opt_prob}. Re-order the queues (permute their labels) 
such that $D_1(\lambda^*_1) \geq D_2(\lambda^*_2) \geq \ldots \geq 
D_N(\lambda^*_N)$. Then, there exist $0=\beta_0\leq \beta_1 \leq \ldots \leq 
\beta_N = \beta_{\max}$ such that $\mbox{supp}(\boldlambda^*_j) \subseteq 
[\beta_{j-1},\beta_j]$ for all $j=1,\ldots,N$. 
\end{corollary}
 
The corollary says that customers are almost segregated by class, i.e., that each 
queue serves a set of customer classes that is nearly disjoint from those served 
in other queues. By nearly disjoint, we mean that the customer classes served at 
distinct queues constitute intervals (closed, open or neither), which may only 
intersect at their boundaries. 
If the measure $\boldeta$ has atoms (e.g., if there are only finitely many classes), 
then it is possible that customers belonging to some of these atoms are split across 
two or more queues. In routing terms, this would imply probabilistic routing to the 
corresponding queues. Secondly, the congestion costs at the queues are ordered 
such that more delay-sensitive customers incur smaller delays. Note that we are 
not claiming that queues with smaller delays have faster servers. Indeed, all servers 
may be identical, or the servers in less congested queues may even be slower! The 
differentiation in congestion costs is an emergent property of the optimal solution 
rather than a consequence of intrinsic differences between servers.
 

Next, we consider the same model, augmented with admission prices. 
Without loss of generality, we take $c_1 < c_2 < \ldots <c_N;$  if 
$c_i=c_j,$ then we can collapse these two queues into a single queue 
whose delay function is the inf-convolution of the delay functions of its 
constituent queues, i.e.,
$$  
D(\lambda)=\inf \{ D_i(\lambda_i)+ D_j(\lambda_j): 
\lambda_i, \lambda_j \ge 0, \lambda_i+\lambda_j=\lambda \}.
$$
Each customer seeks to join a queue that minimizes the sum of the admission price, 
which is common to all classes, and the expected congestion cost, which is weighted 
by its own delay-sensitivity. We wrote down conditions in \eqref{wardrop} for a 
decomposition of the arrival intensity measure $\boldeta$  to be a Wardrop equilibrium. 
We now show that any Wardrop equilibrium has the same structure that we 
demonstrated above for a social optimum.

\begin{theorem} 
\label{thm:wardrop} 
Suppose $(\boldlambda_1^W,\ldots,\boldlambda_N^W)$ satisfies the conditions in 
\eqref{wardrop}, i.e., is a Wardrop equilibrium. Suppose $i$ and $j$ are distinct 
queues, $\beta_2 > \beta_1 \geq 0$, and
$$
\boldlambda^W_j([\beta_2,\infty))>0 \mbox{ and } \boldlambda^W_i([0,\beta_1])> 0.
$$
Then $c_j>c_i$. 
\end{theorem}

The theorem says that if some of the customers served at queue $j$ have 
higher delay sensitivity than some of the customers served at queue $i$, then 
the admission price at queue $j$ must be larger. Whereas the social optimum 
does not use queues whose congestion cost at zero load is too high, a queue
could remain unused in a Wardrop equilibrium either because its congestion cost 
at zero load is too high, or because its admission price is too high, or a combination 
of the two.
The theorem implies that the queues segregate traffic by class as follows:

\begin{corollary} 
\label{cor:wardrop-sort} 
Suppose $(\boldlambda_1^W,\ldots,\boldlambda^W_N)$ satisfy the conditions  
in \eqref{wardrop}, with admission prices $c_1<c_2<\ldots<c_N$. Then, 
there exist $0=\beta_0\leq \beta_1 \leq \ldots \leq \beta_N = \beta_{\max}$ 
such that $\mbox{supp}(\boldlambda^W_j) \subseteq [\beta_{j-1},\beta_j]$ 
for all $j=1,\ldots,N$. 
\end{corollary}

An important difference with the social optimum is that the ordering of queues 
by congestion cost at the social optimum is not obvious \emph{a priori}. Hence, 
we do not know which queue will serve more delay-sensitive customers and which 
will serve less delay sensitive ones. On the other hand, at a Wardrop equilibrium, 
queues which charge a higher admission price (and are not idle) will serve more 
delay-sensitive customes than ones which charge a lower admission price.

\section{Proofs} \label{sec:proofs}

We now present proofs of the various results stated in the previous section.

\begin{proof}\emph{of Lemma~\ref{lem:soc_opt}}. 
It is well-known that the set of sub-probability measures on $\Reals_+$ is compact 
in the weak topology. Hence, so too is the set of measures $\boldlambda$ on 
$\Reals_+$ such that $\lambda \leq \eta$, where $\lambda=\boldlambda(\Reals_+)$,  
and $\eta=\boldeta(\Reals_+)<\infty$. By Tychonoff's theorem, the set 
$
\{ (\boldlambda_1,\ldots,\boldlambda_N): \lambda_i \leq \eta \;\forall \; i=1,\ldots,N \}
$
is compact in the product topology. Next, the map $(\boldlambda_1,\ldots,\boldlambda_N) 
\mapsto \boldlambda_1+\ldots+\boldlambda_N$ is continuous in this topology, and so the 
set $\{ (\boldlambda_1,\ldots,\boldlambda_N):\boldlambda_1+\ldots+\boldlambda_N=\boldeta \}$ 
is closed. As it is a closed subset of a compact set, it is compact.

Let $\beta_{\max}=\sup \{ \mbox{supp}(\boldeta) \}$. Then $\beta_{\max}$ is 
finite by assumption. Hence, the support of $\boldlambda_j$ is also restricted to 
$[0,\beta_{\max}]$ for all $j$, and the maps $\boldlambda_j \mapsto \lambdabar_j$ 
are continuous in the weak topology; so, too, are the maps $\boldlambda_j \mapsto 
\lambda_j$. even without requiring bounded support. Finally, since the optimization 
problem \eqref{welfare_opt_prob} is feasible, we can restrict the minimization to a 
set of $(\boldlambda_1,\ldots,\boldlambda_N)$ on which $\mathcal{U}$ is bounded; 
in particular, each $\lambda_j$ is in the domain of $D_j(\cdot)$. On this set, 
$\mathcal{U}$ is continuous in the product topology. Thus, \eqref{welfare_opt_prob} 
involves the minimization of a continuous function over a compact set. Therefore, 
the minimum is attained.
\end{proof}

\begin{proof}\emph{of Lemma~\ref{lem:wardrop_exist}}. 
The constrained optimization problem \eqref{eq:wardrop_opt} seeks the minimum 
of a continuous function over a compact set; this follows along the same lines as 
the proof of Lemma~\ref{lem:soc_opt}. Hence, a minimizer exists.

Let $\boldlambda^W= (\boldlambda^W_1,\ldots,\boldlambda^W_N)$ be one such 
minimizer. Suppose by way of contradiction that it is not a Wardrop equilibirum, i.e., 
that it does not satisfy \eqref{wardrop}. Then, there exist queues $j$ and $k$ such that 
\begin{equation} \label{eq:contradictw} 
\begin{aligned}
&c_j + \beta D_j(\lambda^W_j) > c_k + \beta D_k(\lambda^W_k) \\
&\quad \mbox{ for some } \beta \in \mbox{supp}(\boldlambda^W_j).
\end{aligned}
\end{equation}

By definition of the support, for any $\delta>0$, there is an $\epsilon>0$ such that 
$\boldlambda^W_j((\beta-\delta,\beta+\delta) = \epsilon$. We now define a new 
decomposition of $\boldeta$ which corresponds to shifting the mass in $(\beta-\delta, 
\beta+\delta)$ from queue $j$ to queue $k$. More formally, denote the restriction 
of a measure $\boldmu$ to a set $A$ by $\boldmu|_A$. Define $\boldmu = 
\boldlambda^W_j|_{(\beta-\delta,\beta+\delta)}$. For $\epsilon\in (0,1)$, define 
$$
\boldnu^{\epsilon}_i = \begin{cases}
\boldlambda^W_i, & i\neq j,k \\
\boldlambda^W_j - \epsilon \boldmu^{\beta,\delta}, & i=j, \\
\boldlambda^W_k + \epsilon \boldmu^{\beta,\delta}, & i=k.
\end{cases}
$$
Clearly, $\boldnu^{\epsilon}_i$, $i=1,\ldots,N$ are non-negative measures 
and decompose $\boldeta$, for any $\epsilon \in (0,1)$. We see from 
\eqref{eq:wardrop_opt} that 
\begin{equation*} 
\begin{aligned}
&\hat{\mathcal{U}} (\boldnu^{\epsilon}) - \hat{\mathcal{U}} (\boldlambda^W) \\
&= \int_{\lambda^W_k}^{\lambda^W_k+\epsilon \mu} D_k(x)dx 
- \int_{\lambda^W_j-\epsilon \mu}^{\lambda^W_j} D_j(x)dx \\
&\qquad + \epsilon \int_{\beta-\delta}^{\beta+\delta} 
\frac{c_k-c_j}{\alpha}d\boldmu(\alpha) \\
&= \Bigl( D_k(\lambda^W_k)-D_j(\lambda^W_j)
+ \frac{c_k-c_j}{\beta} \Bigr) \mu \epsilon + o(\epsilon) + O(\delta \epsilon).
\end{aligned}
\end{equation*}
By \eqref{eq:contradictw}, the quantity in the last line above is negative, for small 
enough $\delta$ and $\epsilon$. This contradicts the optimality of $\boldlambda^W$. 
The lemma is proved by contradiction.
 
\end{proof}

\begin{proof}\emph{of Theorem~\ref{thm:prices_optimal}}. 
The proof is by contradiction. Suppose $\boldlambda=(\boldlambda_1^*,\ldots,
\boldlambda_N^*)$ solves the welfare optimization problem, \eqref{welfare_opt_prob}, 
and that the admission prices $c_j$ are set equal to the corresponding Pigouvian taxes, 
defined in \eqref{eq:pigou_tax}. Suppose that $(\boldlambda_1^*,\ldots,
\boldlambda_N^*)$ do not satisfy \eqref{wardrop}, i.e., are not a Wardrop equilibrium 
for these prices. Then, there exist queues $j$ and $k$ such that 
\begin{equation} \label{eq:contradict} 
\begin{aligned}
&c_j + \beta D_j(\lambda^*_j) > c_k + \beta D_k(\lambda^*_k) \\
&\quad \mbox{ for some } \beta \in \mbox{supp}(\boldlambda^*_j).
\end{aligned}
\end{equation}

By definition of the support, for any $\delta>0$, there is an $\epsilon>0$ such that 
$\boldlambda^*_j((\beta-\delta,\beta+\delta) = \epsilon$. We now define a new 
decomposition of $\boldeta$ which corresponds to shifting the mass in $(\beta-\delta, 
\beta+\delta)$ from queue $j$ to queue $k$. Denoting the restriction 
of a measure $\boldmu$ to a set $A$ by $\boldmu|_A$, we define 
$$
\boldlambda^{\beta,\delta}_i = \begin{cases}
\boldlambda^*_i, & i\neq j,k \\
\boldlambda^*_j - \boldlambda^*_j|_{(\beta-\delta,\beta+\delta)}, & i=j, \\
\boldlambda^*_k + \boldlambda^*_j|_{(\beta-\delta,\beta+\delta)}, & i=k.
\end{cases}
$$
Clearly, $\boldlambda^{\beta,\delta}_i$, $i=1,\ldots,N$ are non-negative 
measures, and decompose $\boldeta$. We see from \eqref{welfare_opt_prob} 
that 
\begin{align*} 
&\mathcal{U} (\boldlambda^{\beta,\delta}) - \mathcal{U} (\boldlambda^*) \\
&=\lambdabar^{\beta,\delta}_j D_j(\lambda^{\beta,\delta}_j) + 
\lambdabar^{\beta,\delta}_k D_k(\lambda^{\beta,\delta}_k) \\
&\qquad -\lambdabar^*_j D_j(\lambda^*_j) - \lambdabar^*_k D_k(\lambda^*_k) \\
&= \Bigl( \lambdabar^*_j-\beta \epsilon+O(\delta \epsilon) \Bigr)
\Bigl( D_j(\lambda^*_j) - \epsilon D'_j(\lambda^*_j)+o(\epsilon) \Bigr) \\
&\qquad - \lambdabar^*_j D_j(\lambda^*_j) \\
& \quad + \Bigl( \lambdabar^*_k+\beta \epsilon+O(\delta \epsilon) \Bigr)
\Bigl( D_k(\lambda^*_k) + \epsilon D'_k(\lambda^*_k)+o(\epsilon) \Bigr) \\
&\qquad - \lambdabar^*_k D_k(\lambda^*_k) \\
&= \epsilon \Bigl( \beta D_k(\lambda^*_k) + \lambdabar^*_k D'_k(\lambda^*_k) 
- \beta D_j(\lambda^*_j) - \lambdabar^*_j D'_j(\lambda^*_j) \Bigr) \\
&\qquad +O(\delta \epsilon) + o(\epsilon).
\end{align*}
Substituting the expression for the Pigouvian taxes $c_j$ and $c_k$ from 
\eqref{eq:pigou_tax} in the above, we get 
\begin{align*}
\mathcal{U} (\boldlambda^{\beta,\delta}) - \mathcal{U} (\boldlambda^*) 
= &\epsilon \bigl( c_k + \beta D_k(\lambda^*_k) - c_j - \beta D_j(\lambda^*_j) \bigr) \\
&+ O(\delta \epsilon) + o(\epsilon).
\end{align*}
If we let $\delta$ decrease to zero, then so does $\epsilon$, and the last two terms 
in the expression above are negligible compared to the first. Hence, it follows from 
the above and \eqref{eq:contradict} that $\mathcal{U} (\boldlambda_1^{\beta,\delta},
\ldots,\boldlambda_N^{\beta,\delta}) - \mathcal{U} (\boldlambda_1^*,\ldots,
\boldlambda_N^*) <0$ for $\delta$ sufficiently small. This contradicts the assumed 
optimality of $(\boldlambda^*_1,\ldots,\boldlambda^*_N)$.

We have thus shown by contradiction that the conditions, \eqref{wardrop}, for a 
Wardrop equilibrium must be satisfied at a socially optimal allocation when the 
admission prices are given by Pigouvian taxes.  
\end{proof}

\begin{proof}\emph{of Lemma~\ref{lem:nonconvex}}.
The proof is an exercise in calculus. The set of 
$(\lambda_1,\lambdabar_1, \lambda_2, \lambdabar_2)$ satisfying the constraints 
in \eqref{welfare_opt_2class} is convex. A necessary condition for the objective function 
to be convex on the feasible set is that the Hessian of $U$ be positive semi-definite on 
the subspace $\{ (x_1,x_2,x_3,x_4): x_1+x_3=0, x_2+x_4=0 \}$ of feasible 
deviations, at each feasible point $(\lambda_1,\lambdabar_1, \lambda_2, \lambdabar_2)$.

Denoting the Hessian by $[D^2U]$, we consider the quadratic form 
\begin{align*}
& (x_1 , x_2 , x_3 , x_4) [D^2 U(\lambda_1, \lambdabar_1, \lambda_2, \lambdabar_2)] 
(x_1, x_2, x_3, x_4)^T \\
&= \frac{2\lambdabar_1 x_1^2}{(1-\lambda_1)^3} + \frac{2x_1 x_2}{(1-\lambda_1)^2}
+ \frac{2\lambdabar_2 x_3^2}{(1-\lambda_2)^3} + \frac{2x_3 x_4}{(1-\lambda_2)^2} \\
&= \Bigl( \frac{2\lambdabar_1}{(1-\lambda_1)^3} + \frac{2\lambdabar_2}{(1-\lambda_2)^3}
 \Bigr) x_1^2 \\
&\qquad + \Bigl( \frac{2}{(1-\lambda_1)^2} + \frac{2}{(1-\lambda_2)^2} \Bigr) x_1 x_2,
\end{align*} 
where we have used the fact that $x_1=-x_3$ and $x_2=-x_4$ on the subspace of interest 
to obtain the second equality. Now, it is is clear that the expression above can be made 
negative by choosing $x_1$ and $x_2$ non-zero and of opposite signs, and $x_1$ sufficiently 
small in absolute value. 

In other words, the quadratic form is not always non-negative, i.e., the Hessian is not 
positive semi-definite on the subspace of interest. Therefore, the objective function $U$ 
is not convex on the feasible set.  
\end{proof}

\begin{proof}\emph{of Theorem~\ref{thm:welfare-opt-cts}}.
Let $\boldlambda^*=(\boldlambda_1^*,\ldots,\boldlambda_N^*)$ solve 
\eqref{welfare_opt_prob}, and let $i$, $j$, $\beta_1$ and $\beta_2$ be as in 
the statement of the theorem. We shall prove the theorem by contradiction.
 
Suppose first that $\lambda_i^*>0$ and that $D_i(\lambda_i^*)<D_j(\lambda_j^*)$. 
We shall show that shifting a small mass of customer from queue $j$ to queue $i$ and 
an equal mass from $i$ to $j$ reduces the social cost, contradicting the optimality of 
$\boldlambda^*$. Let $\boldmu_i$ and $\boldmu_j$ be measures such that 
\begin{align*}
&\boldmu_i \leq \boldlambda_i, \; \boldmu_j \leq \boldlambda_j, \; \mu_i=\mu_j>0, \\
&\mbox{supp}(\boldmu_i)\subseteq [0,\beta_i], \; 
\mbox{supp}(\boldmu_j)\subseteq [\beta_j,\infty). 
\end{align*}
It is clear from the assumptions that such measures exist. Since $\beta_j>\beta_i$, 
we also have $\mubar_j>\mubar_i$.

Consider the measures $\tilde \boldlambda$ defined as follows:
$$
\tilde{\boldlambda}_k = \begin{cases}
\boldlambda^*_k, & k\neq i,j, \\
\boldlambda^*_i + \boldmu_j - \boldmu_i, & k=i, \\
\boldlambda^*_j - \boldmu_j + \boldmu_i, & k=j.
\end{cases}
$$
Then, $\tilde{\lambda}_k=\lambda^*_k$ for all $k$, since equal masses 
are swapped between queues $i$ and $j$ while flows into all other queues 
are unchanged. Hence, the congestion costs $D_k$ at all queues remain 
unchanged. Thus, we get 
$$
\mathcal{U}(\tilde{\boldlambda})-\mathcal{U}(\boldlambda^*) 
= (\mubar_j-\mubar_i)\Bigl( D_i(\lambda^*_i) - D_j(\lambda^*_j) \Bigr)<0,
$$
since $\mubar_j>\mubar_i$ as noted, while $D_i(\lambda_i^*)<D_j(\lambda_j^*)$ 
by assumption. But this contradicts the optimality of $\boldlambda^*$. Thus, we 
cannot have $D_i(\lambda_i^*)<D_j(\lambda_j^*)$ and $\lambda^*_i>0$.

Suppose next that $\lambda^*_i>0$ and $D_i(\lambda_i^*)=D_j(\lambda^*_j)$. 
Let $\tilde{\boldlambda}$ be as above, and define 
$$
\boldlambda^{\alpha}=\alpha \tilde{\boldlambda}+(1-\alpha)\boldlambda^*, 
\quad \alpha \in [0,1].
$$
Then, for all $\alpha \in [0,1]$, $\lambda^{\alpha}_i=\lambda^*_i$ and 
$\lambda^{\alpha}_j=\lambda^*_j$, so $D_i(\lambda^{\alpha}_i) = 
D_i(\lambda^*_i) = D_j(\lambda^*_j) =D_j(\lambda^{\alpha}_j)$. Hence, 
$\mathcal{U}(\boldlambda^{\alpha})=\mathcal{U}(\boldlambda^*)$, which 
implies that $(\boldlambda^{\alpha}_1,\ldots,\boldlambda^{\alpha}_N)$ solve the 
welfare optimization problem, \eqref{welfare_opt_prob}, for every $\alpha \in [0,1]$.

Now, for $\alpha \in (0,1)$, and small enough $|\epsilon|$, define the measures 
$\boldnu^{\alpha,\epsilon}_k$, $k=1,\ldots,N$, by 
$$
\boldnu^{\alpha,\epsilon}_k = \begin{cases}
\boldlambda^{\alpha}_k, & k\neq i,j, \\
\boldlambda^{\alpha}_i + \epsilon \boldmu_j, & k=i, \\
\boldlambda^{\alpha}_j - \epsilon \boldmu_j, & k=j.
\end{cases}
$$
If $|\epsilon|$ is sufficiently small, depending on $\alpha$, then these are 
non-negative measures. We now have 
\begin{align*}
&\mathcal{U}(\boldnu^{\alpha,\epsilon})-\mathcal{U}(\boldlambda^{\alpha}) \\
&= \nubar^{\alpha,\epsilon}_i D_i(\nu^{\alpha,\epsilon}_i) +
\nubar^{\alpha,\epsilon}_j D_j(\nu^{\alpha,\epsilon}_j) \\
&\qquad -\lambdabar^{\alpha}_i D_i(\lambda^{\alpha}_i) - 
\lambdabar^{\alpha}_j D_j(\lambda^{\alpha}_j) \\
&= \epsilon \Bigl( \mubar_j D_i(\lambda^{\alpha}_i) + 
\mu_j \lambdabar^{\alpha}_i D'_i(\lambda^{\alpha}_i) \Bigr. \\
&\qquad \Bigl. - \mubar_j D_j(\lambda^{\alpha}_j) 
- \mu_j \lambdabar^{\alpha}_j D'_j(\lambda^{\alpha}_j) \Bigr) + o(\epsilon).
\end{align*}
For $\mathcal{U}(\boldlambda^{\alpha})$ to be a global minimum, the coefficient 
of $\epsilon$ in the above expression must be zero. Thus, 
$$
\mu_j \Bigl( \lambdabar^{\alpha}_i D'_i(\lambda^{\alpha}_i) - 
\lambdabar^{\alpha}_j D'_j(\lambda^{\alpha}_j) \Bigr) =
\mubar_j (D_j(\lambda^{\alpha}_j)-D_i(\lambda^{\alpha}_i)).
$$
But $\lambda^{\alpha}_k=\lambda^*_k$ for all $\alpha\in [0,1]$ and $k=1,\ldots,N$. 
Combining this with the fact that $D_i(\lambda^*_i)=D_j(\lambda^*_j)$ by assumption, 
we can rewrite the last equation as
\begin{equation} \label{eq:kkt_alpha} 
\mu_j \Bigl( \lambdabar^{\alpha}_i D'_i(\lambda^*_i) - 
\lambdabar^{\alpha}_j D'_j(\lambda^*_j) \Bigr) = 0.
\end{equation}
Now, $\lambdabar^{\alpha}_i$ is strictly increasing in $\alpha$ and 
$\lambdabar^{\alpha}_j$ is strictly decreasing, as $\boldlambda^{\alpha}$ is 
obtained by swapping a volume of more delay-sensitive traffic in queue $j$ for 
an equal volume of less delay-sensitive traffic in queue $i$, and these volumes 
are increasing in $\alpha$. Moreover, $D'_i(\lambda^*_i)$ and 
$D'_j(\lambda^*_j)$ are strictly positive, and hence non-zero, by assumption. 
It follows that \eqref{eq:kkt_alpha} cannot hold for all $\alpha \in (0,1)$, or 
even for two distinct values of $\alpha$.

Thus, we have shown by contradiction that we cannot have $\lambda^*_i>0$ and 
$D_i(\lambda^*_i)=D_j(\lambda^*_j)$. It only remains to consider the possibility 
that $\lambda^*_i=0$. Let $\boldmu_j$ be as above. Fix $\epsilon>0$ sufficiently 
small, and define the measures $\boldnu^{\epsilon}$ as follows:
$$
\boldnu^{\epsilon}_k = \begin{cases}
\boldlambda^*_k, & k\neq i,j, \\
\epsilon \boldmu_j, & k=i, \\
\boldlambda^*_j - \epsilon \boldmu_j, & k=j.
\end{cases}
$$
Then, we have 
$$
\mathcal{U}(\boldnu)-\mathcal{U}(\boldlambda^*) = 
\epsilon \mu_j (D_i(0)-D_j(\lambda^*_j)) - 
\epsilon \mu_j \lambdabar^*_j D'_j(\lambda^*_j) + o(\epsilon).
$$
Since $D'_j(\lambda^*_j)>0$, the above quantity is negative, 
contradicting the optimality of $\boldlambda^*$, unless 
$D_i(0)>D_j(\lambda^*_j)$. This completes the proof of the theorem.  
\end{proof}

\begin{proof}\emph{of Corollary~\ref{cor:sorting-cts}}.
Suppose the corollary is false. Then, there is a solution $\boldlambda^*=
(\boldlambda^*_1,\ldots,\boldlambda^*_N)$ of \eqref{welfare_opt_prob}, and 
queues $i$ and $j$, such that $D_i(\lambda^*_i) \geq D_j(\lambda^*_j)$ but 
queue $i$ also serves a non-zero mass of customers who are more delay-sensitive 
than some of the customers served in queue $j$. More precisely, there exist 
$\beta_2>\beta_1$ such that $\boldlambda^*_i([\beta_2,\infty))>0$ and 
$\boldlambda^*_j([0,\beta_1])>0$. But this contradicts 
Theorem~\ref{thm:welfare-opt-cts}.  
\end{proof}

\begin{proof}\emph{of Theorem~\ref{thm:wardrop}}.
Suppose $\boldlambda^W=(\boldlambda_1^W,\ldots,\boldlambda_N^W)$ satisfies the 
conditions in \eqref{wardrop}. Suppose $i$ and $j$ are distinct queues and $\beta_2 
> \beta_1 \geq 0$ are such that 
$$
\boldlambda^W_j([\beta_2,\infty))>0 \mbox{ and } \boldlambda^W_i([0,\beta_1])>0.
$$
Pick $\beta \leq \beta_1 \in \mbox{supp}(\boldlambda^W_i)$ and $\gamma \geq 
\beta_2 \in \mbox{supp}(\boldlambda^W_j)$. We have by \eqref{wardrop} that 
\begin{equation} \label{wardrop_ineqs}
\begin{aligned}
c_i + \beta D_i(\lambda^W_i) &\leq c_j + \beta D_j(\lambda^W_j), \\
c_j + \gamma D_j(\lambda^W_j) &\leq c_i + \gamma D_i(\lambda^W_i).
\end{aligned} 
\end{equation}
It follows from these inequalities that $(\gamma-\beta)(D_i(\lambda^W_i)
-D_j(\lambda^W_j) \leq 0$. Since $\gamma>\beta$, it follows that $D_i(\lambda^W_i) 
\geq D_j(\lambda^W_j)$. Substituting this in \eqref{wardrop_ineqs}, we obtain that 
$c_i \leq c_j$. As it was assumed that admission prices are all distinct, we have $c_i>c_j$, 
as claimed.
\end{proof}

\begin{proof}\emph{of Corollary~\ref{cor:wardrop-sort}}. 
Consider two queues $i$ and $j$. Suppose $\beta_1 \in \mbox{supp}(\boldlambda^W_i)$, 
$\beta_2 \in \mbox{supp}(\boldlambda^W_j)$ and $\beta_1<\beta_2$. Then, there is a 
$\delta>0$ sufficiently small that 
\begin{align*}
&\boldlambda^W_j([\beta_2-\delta,\infty))>0, \quad \boldlambda^W_i([0,\beta_1+\delta])>0, \\
&\beta-\delta>\beta_1+\delta.
\end{align*}
Hence, by Theorem~\ref{thm:wardrop}, $c_j>c_i$, i.e., $j>i$. This proves the corollary.
\end{proof}

\RemoveThis{
  Thus
  \begin{eqnarray*}
    \hat{U}(P^{\prime}) &=& \sum_{i=1}^M \frac{\lambda_i}{\lambda} 
    \psi\left(\sum_{i=1}^M \frac{\lambda_i}{\lambda}(U_i(P))\right).
  \end{eqnarray*}
  Note that $P^+$ is more fair compared to $P$ since $U_i(P^+)$ is the
  same for the different classes (refer to \eqref{eq:fair}) which may
  not be the case for the allocation $P$.  Clearly, for a more fair
  allocation $P^+$ we have $\hat{U}(P) \geq \hat{U}(P^+).$ From
  Jensen's ineqality, more unfair allocations are more heavily
  penalized by the expectation of the resulting cost function.
  %
  %
  %
}

\section{Summary and Discussion}
\label{sec:conclusion}
We considered a very general model of multiple parallel queues serving a 
heterogeneous customer population, and studied the problem of routing 
customers to queues so as to maximize social welfare. We characterized 
certain structural properties of the welfare-optimizing allocation. We also 
considered selfish routing decisions made by individual customers when 
the queues charge admission prices, and characterized the structure of 
Wardrop equilibria. Finally, we showed that, if the admission prices at the 
queues are set equal to the congestion externalities at a socially optimal 
allocation, then the social optimum coincides with a Wardrop equilibrium. 

The setting we studied was very general, and encompassed a variety 
of applications with congestion externalities. Nevertheless, some of the 
assumptions are restrictive. We model customer heterogeneity by 
applying different multipliers to a common measure of congestion cost 
at each queue. But it might be the case that some customers care about 
mean delay, while others care about the probability of exceeding a certain 
threshold. In that case, no multiplier on the congestion cost would be 
appropriate for capturing this diversity. Another restrictive assumption 
is that customers may differ in delay sensitivity, but not in the distribution 
of the workload they bring into the system. Indeed, this is why Pigouvian 
tolls depend on the queue, but not on the customer class. If this 
assumption were relaxed, the externality imposed by a customer would 
depend on its workload, and hence on its class; this would need to be taken 
into account in setting Pigouvian tolls.

We briefly discussed the difficulty of determining the optimal allocation. 
We showed that the optimization problem is non-convex, but did not 
prove that it is hard. The structural properties of the optimal allocation 
that we established do not resolve this question, as the optimal ordering 
of the queues is unknown. Even if the optimal ordering were given, it is 
not entirely obvious that the thresholds can be computed efficiently. 
Likewise, the computational complexity of determining the Wardrop 
equilibria is also unknown. Note that the ordering of queues in this case 
is determined by the given prices. Thus, one open problem for future 
research is developing efficient algorithms for these problems, or proving 
that they are hard.

A second question concerns the informational constraints on the model.
We have assumed that the arrival intensity measure is known, and 
available as input to determining a socially optimal allocation or setting 
admission prices. In practice, this information is unlikely to be available, 
but needs to be inferred from observation. If a customer's delay sensitivity 
is revealed upon arrival, then the arrival distribution can easily be measured. 
But eliciting delay sensitivities truthfully can be a challenge in practice. 
It is an open question whether it is still possible to set admission prices 
in such a way as to ensure that the Wardrop equilibrium either coincides with 
the welfare optimizing allocation,or approximates it to within some factor.

Finally, we have assumed that a benevolent mechanism designer sets
admission prices to maximize social welfare; it is interesting to ask
what happens if the admission prices are set by a revenue maximizing
service provider. Further, in such a revenue maximizing scenario it
would be interesting to see if competing service providers can sustain
differentiated services.

\section{Acknowledgements}
This work was done while the first author was affiliated with IIT Bombay, India.
Bodas and Manjunath acknowledge support from the 
Bharti Centre for Communication at IIT Bombay, CEFIPRA and IFCAM.

\bibliography{pricing-queues}

\bibliographystyle{plain}

\end{document}